\documentstyle[prl,twocolumn,aps,psfig]{revtex}
\begin{document}
\draft
\title{Reaction--Diffusion Fronts under Stochastic Advection}
\author{A.C. Mart\'\i ,$^{1}$ 
F. Sagu\'es$^{2}$ and J.M. Sancho$^{1}$}

\address{
$^1$ Departament d'Estructura i Constituents de la Mat\`eria,\\
Universitat de Barcelona,\\
Av. Diagonal 647,\\
E-08028 Barcelona, Spain.\\
$^2$ Departament de Qu\'{\i}mica F\'{\i}sica,\\
Universitat de Barcelona,\\
Av. Diagonal 647,\\
E-08028 Barcelona, Spain.\\
}
\date{\today}
\maketitle

\begin{abstract}

We study front propagation in stirred media using a simplified
modelization of the turbulent flow.  Computer simulations reveal the
existence of the two limiting propagation modes observed in recent
experiments with liquid phase isothermal reactions. These two modes
respectively correspond to a wrinkled although sharp propagating interface
and to a broadened one.  Specific laws relative to the enhancement of the
front velocity in each regime are confirmed by our simulations.

\end{abstract}


\vskip5mm

\section{Introduction}

Front propagation problems have been receiving an increasing attention
due to its relevance for non--equilibrium studies in a wide variety of
fields \cite{cross}. Most of these studies are concerned with the
velocity selection problem in deterministic systems or homogeneous
medium (reaction--diffusion equations) \cite{vansaarlos}. Very recently
front propagation in modulated \cite{joan1} or stochastic media has
been discussed \cite{joan2}.

Self sustained chemical fronts propagating in quiescent media advance
steadily at a velocity $v_0$ which results purely from the interplay
between the chemical time scale $\tau _{chem}$ and the species
diffusivity coefficient $D$: $v_0\sim (D \tau
^{-1}_{chem})^{1/2}$. However, in stirred media, additional length and
time scales come to play a fundamental role into such a dynamical
process. The interface, either wrinkled or broadened by the turbulent
flow, propagates at a velocity $v_T$ which is enhanced relative to
$v_0$ to an extent which depends not only on the invested energy but
also on the coupling between the intrinsic length and time scales and
those externally introduced by the stirring mechanism
\cite{Ronney95a}.

Needless to say, the addressed problem is of singular importance either
as a way to inquiry about fundamental aspects of turbulent flows, but
mainly for its practical relevance to combustion processes
\cite{Williams85}. However, abundant studies in either way motivated
have not been able yet to resolve the existing considerable
discrepancies between theoretical predictions and experimental
results. These discrepancies concern fundamental issues such a
turbulent propagation rates, velocity quenching effects, role of
turbulent spectra, etc.  \cite{Ronney95a}.

From the experimental point of view much has been gained recently by
enlarging the scope of the problem when going from combustion
processes to liquid phase reactions \cite{Ronney95}. In this way
experimentalists more confidently approach the set of simplifying
assumptions invoked by most of the theoretical models. These later
ones, either based on the so--called G--equation \cite{Kernstein88},
passage--rate analysis \cite{Kernstein92} or scale--covariance laws
\cite{Pocheau96} tend to focus on a single mode of turbulent
propagation i.e. that identified later on as based on the so--called
mechanism. On the other hand, when resorting to computing simulations
\cite{Haworth92}, numerical accuracy commonly restricts the examined
scenario to small stirring intensities.  On the whole, it is then
likely to expect a future demand for more versatile computer
approaches to compare with more general and better controlled
experimental observations.

Advancing towards that direction, we report here on our computer
results for interfaces propagating in stirred media.  Stirring is
mimicked by using an stochastic differential equation (SDE) based
algorithm to simulate random, stationary, isotropic and statistically
homogeneous flows with well prescribed properties \cite{marti}. In
spite of this ``ad hoc'' procedure, the pair of experimentally
identified limiting modes of front propagation in liquids phases
\cite{Ronney95}, i.e. the aforementioned Huygens propagation (HP) and
the distributed reaction zone (DRZ) regimes, are clearly exhibited in
our simulations. In addition we do not suffer from strict limitations
concerning stirring intensity, what enables us to look for broad range
scalings of the turbulent propagation rates. Finally, the main
advantage of our model is the easy control we gain over the flow
parameters more relevant for front propagation in stirred media i.e.,
the intensity $u_0^2$, spectra $E(k)$, spatial correlation length 
$l_0 $, and time correlation $t_0$.  In particular on what follows we
adopt the Kraichnan spectrum \cite{Kraichnan70}, describing a widely
distributed band of excitations around a peak centered at some
well--defined wavenumber $k_0$ At this point is worth emphasizing that
the proposed methodology is totally adapted to a statistical
description of turbulence \cite{McComb95} and belongs to the class of
the so--called synthetic turbulence generating models \cite{Juneja94}.

The organization of this paper is as follows. The
reaction--diffusion--advection model and the 
existing analytical prediction on front propagation in 
turbulent media are briefly reviewed in Sec. \ref{sec:2}.
Our numerical results are presented and commented according the above
theoretical predictions in Sec.~\ref{sec:3}.

\section{Turbulent Front Model and Analytical Predictions}
\label{sec:2}
A detailed presentation of the algorithm can be found in \cite{marti}, but 
it is worth summarizing here its main ingredients:

i) A stream function $\eta ({\bf r},t)$ is introduced trough a
SDE written
in terms of a zero mean Gaussian white noise 
$\mbox{\boldmath $ \zeta $}({\bf r},t)$
\begin{equation}
\label{langevin}{\frac{\partial \eta ({\bf r},t)}{{\partial t}}}=\nu \nabla
^2\eta ({\bf r},t) +Q[\lambda^2 \nabla^2] \nabla . \mbox{\boldmath $
\zeta $} ({\bf r},t),
\end{equation}
\begin{equation}
\label{ruido}\left\langle \zeta ^i({\bf r}_1,t_1)\zeta ^j({\bf r}%
_2,t_2)\right\rangle =2\epsilon _0\nu \delta (t_1-t_2)\delta ({\bf r}_1-{\bf %
r}_2)\delta ^{ij}.
\end{equation}

ii)From the stochastic field $\eta $, the 2--d incompressible advecting flow
is obtained  as usual
\begin{equation}
\label{defvel}
{\bf v }({\bf r},t) = \left( -{\frac{{\partial \eta ({\bf r},t)} }{{%
\partial y}}},{\frac{{\ \partial \eta ({\bf r},t)} }{{\ \partial x
}}}\right).
\end{equation}
The two dimensional space is chosen here for several reasons.  First
of all, it is chosen  for the sake of simplicity of numerical simulations.  
Secondly
because the essential trends of the phenomenology we want to study are
independent of the spatial dimension. Recent experiments in chemical
reactions in quasi 2-d geometries \cite{Ronney95} and theoretical
analysis \cite{Pocheau96} support this assumption.

iii) The whole procedure is Fourier transformed and accordingly discretized.

iv) Within this scheme, the homogeneous, isotropic and stationary
velocity correlation $R(r,s)$ and the energy spectrum $E(k)$ can be
properly defined and easily evaluated \cite{marti}.  From these
quantities, the stirring intensity $u_0^2$, and the integral time
$t_0$ and length $l_0$ scales of the flow can be obtained using the
following standard prescription:
\begin{eqnarray}
u_0^{\ 2}    & =   &  R(0,0)= \int_0^{\infty} dk E(k), \nonumber \\
t_0 &  = &  {1 \over { u_0^{\ 2}}} \int_0^{\infty } ds \  R(0,s),\\
\label{physpar}
l_0 & = & {1 \over { u_0^{\ 2}}} \int_0^{\infty } dr \ R(r,0).
\nonumber
\end{eqnarray}

The  adopted Kraichnan spectrum has the following form
\cite{Kraichnan70}
\begin{equation}
\label{Kraichnan}
E(k)\propto k^3\exp \left[ -{ \lambda^2 k^2}\right].
\end{equation}
This energy distribution is reproduced by prescribing
\begin{equation} Q(\lambda ^2\nabla
^2)=\exp \left( {\frac{\lambda ^2\nabla ^2}{{2}}}\right) ,
\end{equation}
On the other hand, the basic statistical parameters of the flow
introduced above expressed in terms of the input parameters, $\epsilon
_0, \lambda, \nu $, read
\begin{equation}
u_0^2=\frac{\epsilon _0}{8\pi \lambda ^4};  \,\,\,\,\,
t_0=\frac{\lambda ^2}\nu ; \,\,\,\,\,
l_0=\frac{\lambda \sqrt{\pi }}2.
\end{equation}

The next step is the formulation of the reaction--diffusion--advection
scheme.  We model this situation by means of a dynamical equation for
the passive scalar field $\psi ({\bf r},t)$ in a two dimensional
space:
\begin{equation}
{\frac{\partial \psi }{{\partial t}}}=D\nabla ^2\psi + f(\psi) - 
\nabla .({\bf v}\psi ).
\label{scalar}
\end{equation}
where $f(\psi)$ is a nonlinear reaction term with at least two steady
states. In our simulation we have taken $f(\psi)=\psi^2- \psi^3$. Our
results are independent of the particular form of $f(\psi)$ chosen in
the simulations.

In the absence of stirring (${\bf v}=0$), a stable planar front
propagates the stable state $\psi =1$ (``products'') into the invaded
unstable one $\psi =0$ (``reactants''). The dimensionless front
velocity and front thickness respectively read:
\begin{equation}
v_0=\sqrt{\frac{D}{2
}},\,\,\,\,\,\,\,\, \delta_0= \sqrt{ 8 D}.
\label{fund}
\end{equation}

Superposing stirring, two limiting regimes of front propagation are
found in our simulations. First when the typical length scale of the
flow $l_0$ is larger than the intrinsic one associated to the
reaction--diffusion dynamics $\delta _0$, we observe a distorted front
which propagates as a still rather sharp interface. Such a propagation
mechanism is known in the combustion literature as the ``thin flame,''
``flamelet'' or ``reaction sheet'' regime \cite{Ronney95}.  Contrarily
when $l_0$ is smaller than $\delta _0$, we observe what is referred
in the literature as a ``DRZ'' regime \cite{Ronney95} i.e., a broadened
front disrupted by the stirring flow. In both cases, turbulent
propagation rates are larger than in quiescent media (parameter values
of the presented simulations are quoted in the figure captions).

Each one of the previously identified conditions correspond to a
specific propagation mechanism. The common rationale behind the ``thin
flame'' mode is based on a HP--like argument: the front has the same
local structure as in the planar case with normal velocity given by
$v_0$, but its length increases due to wrinkling. This results on
faster propagation velocities, in such a way that the relative
increment in the velocity is equal to the relative increment of the
length. On the other hand stirring is assumed to affect the velocity
in the DRZ regime by solely increasing diffusive transport inside the
broadened front. Note in passing that in neither case any effect of
the stirring on the intrinsic chemical time scale is considered.

Before to translate these arguments into quantitative terms let us
define precisely the quantities involved. $L_0$ will be the length of
the planar front which coincides with the lateral length of the
system. $L_T$ is the length of the curved front evaluated numerically
as the length of the curve level of the field at the value $\psi =
1/2$. $D_T$ is the effective diffusion when turbulent flow is
present. It is numerically evaluated by simulating the diffusion of a
passive scalar under the influence of the same flow free of any
chemical front.   
%Thus the determination of $D_T$ is independent
%of the front dynamics. 
The velocity of the front under the turbulent flow $v_T$ is evaluated
numerically as the increment of the products per unit of time and
normalized to $L_0$.  

According to our previous arguments we can
establish the following two analytical results:

i) In the HP regime,
\begin{equation}
\frac{L_T^{}}{L_0}=\frac{v_T^{}}{v_0}\equiv S\stackrel{}{}.
\label{HP:eq}
\end{equation}

ii) Contrarily, in the DRZ mode, we simply adapt the first fundamental
relation of Eq. (\ref{fund}) to obtain
\begin{equation}
S=\frac{v_T}{v_0}=\left( \frac{D_T}D\right) ^{\frac 12}.
\label{DRZ:eq}
\end{equation}

\section{Numerical Results and Comments}
\label{sec:3}
The next and most involved step consists, however, in using
Eqs.~(\ref{HP:eq}) and (\ref{DRZ:eq}) to compare $v_T$, or its 
dimensionless form $S$, as a function of the stirring
intensity $u_0^2$, or its dimensionless value 
$\frac {u_0}{v_0}\equiv Q$ with the results obtained from the
numerical simulations of the Eq. (8).
Let's consider separately on what follows the HP and DRZ regimes.

\subsection{HP mode}

On what refers to the HP mode, our first task was to check relation
(\ref{HP:eq}).  The collected data for the different values of $u_0^2$
are summarized in Fig.~1. For the sake of comparison, we include in
this figure results obtained from two additional and somewhat related
stirring conditions. The first one, hereafter referred as frozen
stirring, corresponds to a fixed configuration of the random flow. The
second one, referred on what follows as periodic stirring, represents
nothing but the limit of a deterministic and single scale flow,
constructed from the single mode stream function
\begin{equation}
\eta (x,y)= \eta _0 \cos ( \frac{n \pi x}{L}) \cos(\frac{n\pi y}{L}),
\end{equation}
representing a periodic array of $n \times n$ eddies, where $L$ is the
system size.  According to Fig.~1, the geometric argument leading to
(\ref{HP:eq}) seems well--supported by our simulations with the simple
exception of those situations involving very intense periodic flows,
were the presence of overhangs (leading to the formation of isolated
islands of reactants) is unavoidable. Note in this respect that the
time evolution of those islands contribute positively to the computed
velocity, measured as the time variation of the rate of occupation of
the $\psi =1$ state, but negatively to the front length.

\subsection{DRZ mode}

Results for the DRZ propagating mode are summarized in Fig.~2.  Here
again results for $S$ obtained by directly simulating the front
propagation dynamics under the three different stirring mechanisms so
far considered are plotted together.  Computer results are compared
with the corresponding theoretically predicted values based on
Eq.~(\ref{DRZ:eq}) above.  The agreement based on Eq.~(\ref{DRZ:eq})
is remarkable for the whole range of $Q$ values here considered and
irrespective of the stirring flow.  Explicit theoretical dependences
of $D_T$ on $u_0^2$ valid respectively in two asymptotic cases are:

i) For a random flow in the weak stirring limit and small integral time $t_0$
\begin{equation}
D_T - D =  
u_0^2 t_0
\end{equation}
A result obtained from Eq. (8) considering ${\bf v}$ as a Gaussian
white noise \cite{marti} and,

ii) For a periodic flows in the limit of
small Peclet number \cite{Moffat},
\begin{equation}
D_T - D \sim  u_0^2 /D. 
\end{equation}
These two analytical results used in relation to Eq. (\ref{DRZ:eq})
above lead to the theoretical estimates also shown in Fig.~2
(continuous lines).

A last remark concerning these two figures can clarify these
results. The argument leading to Eq. (10) and Fig.~1 is pure kinematical
or geometrical and only depends on the fact that the front is narrow
enough. Thus different dynamical situations fit the prediction once
this condition is fulfilled. Contrarily the argument for the DRZ
(Fig.~2) case is dynamical so we observe different behaviors for the
three different flows.

In summary, a recently proposed algorithm to model random stirring has
been used in relation with front propagation. Although artificial in
nature such a model reproduces realistic scenarios of front
propagation modes.  These conclusions together with the versatile
computer implementation of the proposed algorithm give us new hopes to
address some of the open issues in the field of turbulent front
propagation.  Research in this direction basically aimed at finding
the power law dependences of the propagating velocities on the
stirring intensities and their fits to some of the most well--known
relations proposed in the literature, are being presently conducted
and will be reported elsewhere.

%-----------------------------------------------------------------------
\begin{acknowledgements}

We acknowledge financial support by Comision Interministerial de
Ciencia y Tecnologia, (Projects, PB93-0769, PB93-0759) and Centre de
Supercomputaci\'o de Catalunya, Comissionat per Universitats i Recerca
de la Generalitat de Catalunya. A.C.M. also acknowledges partial
support from the CONICYT (Uruguay) and the Programa Mutis (ICI,
Spain). We also acknowledge Prof. Ronney for providing us with
experimental data on turbulent chemical fronts.

\end{acknowledgements}

\begin{figure}
\centerline{\psfig{figure=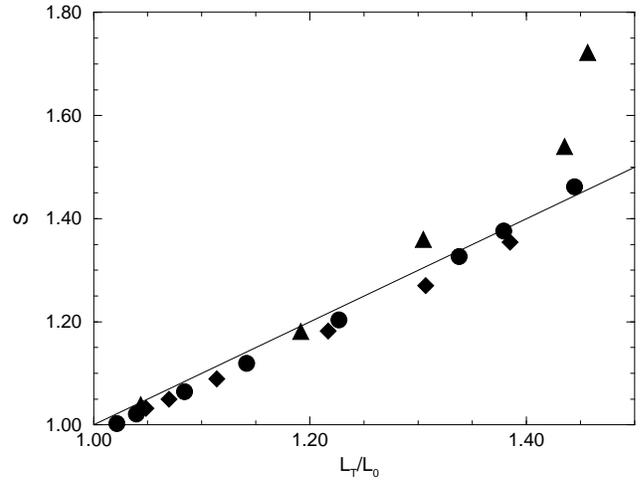,height=7cm}}
\caption{ $S$ versus $L_T / L_0$ (Eq.~10) for turbulent flow
(circles, $l_0=4.0$, $t_0=1.0$), random frozen (squares, $l_0=4.0$)
and periodic eddies (triangles, $8 \times 8$). In all the simulations we
have employed a square lattice of $128 \times 128$ points and unit
spacing $\Delta r =0.5$ and $D=0.3$.}
\label{fig1}
\end{figure}

\begin{figure}
\centerline{\psfig{figure=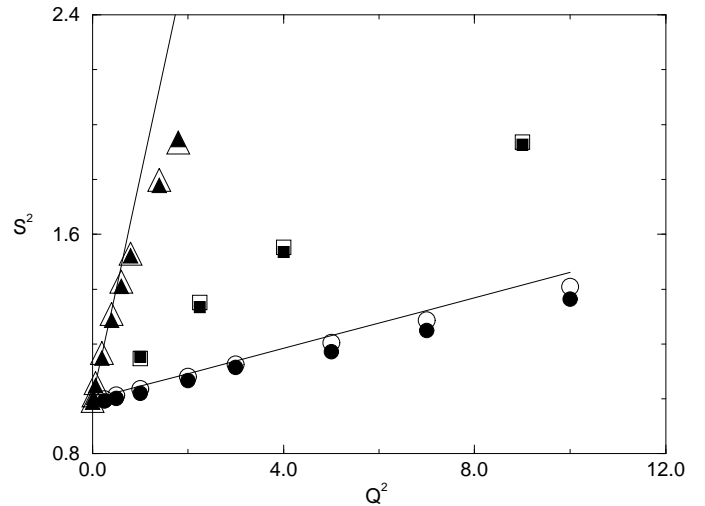,height=7cm}}
\caption{$S^2$ versus $Q^2$ for the DRZ regime.  Full symbols are
simulation results of the front propagation while open symbols stand
for the theoretical Eq. (11) using the effective turbulent diffusion
(see text). Continuous lines correspond to the analytical predictions
based on the perturbative results in (13) and (14). 
Circles, squares, and triangles
correspond to turbulent flow ($l_0=2.0$, $t_0=0.1$), random frozen
($l_0=2.0$) flows and periodic eddies respectively ($16 \times 16 $
eddies) ($D=2.0$).}
\label{fig2}
\end{figure}

\end{document}